\documentstyle [aps,multicol,prl]{revtex}
\begin{document}                                                    
\draft                                                      
\title{ Small
 Magnetic Polaron Picture of 
 Colossal Magnetoresistance in
 Manganites}
\author{Sudhakar Yarlagadda }
\address{Saha Institute of Nuclear Physics, Calcutta, India}             
\date{\today}
\maketitle

\begin{abstract}
 We present
 a small-but-sizeable magnetic  polaron picture where 
transport at high temperatures is activated
 while at low temperatures it is band-like.
 We show that
 both double exchange
and finite bandwidth
effects are important to understand
 colossal magnetoresistance
as well as the coincidence of the
metal-insulator and the ferromagnetic transitions in manganites.
The magnetic transition is 
explained using band-like motion of the polarons.
\end{abstract}

\pacs{PACS numbers: 71.30.+h, 71.38.+i, 72.20.My, 75.10.-b  } 
                                                            

\begin{multicols}{2}
                                                                                

Studies on perovskite manganites
 of the form $\rm A_{1-\delta}B_{\delta} MnO_3$
 (A=La, Pr, Nd, etc.; B=Sr, Ca, Ba, etc.)
have yielded a variety of rich phenomena
as a function of doping $\delta$--coexisting orbital ordering and layered
antiferromagnetism
 at low doping, simultaneous metal-insulator (MI)
 and paramagnetic-ferromagnetic
transitions at intermediate doping ($\delta \sim 0.2-0.4$),
 and charge ordering at higher
doping ($\delta \sim$ 0.5)
 \cite{ramirez}.
To explain the magnetic ordering, de Gennes \cite{degennes} some time
ago had proposed double exchange
mechanism wherein, on account of strong Hund's coupling
between the spin of a mobile hole and the spin of the localized electrons,
 the hopping integral of the itinerant hole
is reduced by half of the cosine of the angle between the
$3/2$ spins of the localized electrons on neighboring sites.
 However it was recognized
by Millis et al. \cite{millis1} that this mechanism itself is not sufficient
 to explain colossal magnetoresistance (CMR). Millis and co-workers have
proposed a model \cite{millis2} which uses Jahn-Teller coupling
 between electrons
and nuclei. However this model does not seem to yield satisfactory results
away from half-filling. R\"{o}der et al. have also emphasized the
importance of Jahn-Teller coupling
 in understanding these manganites \cite{roder}.
Earlier on De Teresa et al. \cite{deteresa}
 have reported evidence for sizeable magnetic polarons
above the ferromagnetic transition temperature.
Recently worledge et al. \cite{worledge} have demonstrated that
their high temperature resistivity data fits well to an adiabatic small
polaron model. All in all there is growing evidence for a 
small magnetic polaron picture to explain CMR.

In this paper we study CMR 
phenomena in perovskite manganites (for $\delta \sim 0.2 - 0.4$)
by considering the 
carriers as adiabatic small-but-sizeable magnetic polarons whose
 high temperature 
 behavior is hopping type and low temperature behavior 
is metal-like.
Our notion of small-but-sizeable polaron is similar to that of a
nearly-small polaron introduced by Eagles \cite{eagles}.
Our treatment builds on the work
of Gosar \cite{gosar},
who included finite band width effects 
by actually
considering a 
small-but-sizeable 
polaron whose wavefunction extends
to the nearest neighbors.
Due to Hund's coupling the electrons on the
nearest neighbors also have the same polarization.
 Our model includes effects due to electron-phonon coupling, on-site Hund's
coupling between itinerant holes and localized electrons, coupling
between nearest neighbor localized spins, and strong on-site
repulsion between two itinerant holes. To understand
MI transition we simplify the Hamiltonian by
accounting for the Hund's coupling through the double exchange
hopping term \cite{degennes}. We show that in 
the absence of a magnetic field
the double exchange and finite bandwidth effects can make
the MI transition coincide with the magnetic transition.
We also find that
in the presence of a magnetic field
 both double exchange 
and finite
band width effects 
lower the resistivity and shift its peak to higher
temperatures and thus can lead to CMR.

      Within a mean-field approach,
  we calculate the magnetization ($M$)
of the localized spins as well as the polarization ($\Delta n$)
of the itinerant hole spins. The magnetization 
$M$ is a function of
 the external magnetic field  and, due to the
 strong Hund's coupling, of also the polarization $\Delta n$.
The polarization $\Delta n$ is in turn a function of 
$M$ and the external field. Thus the two ($M$ and $\Delta n$) are coupled
and have to be solved simultaneously for a given density of holes.
We have studied the magnetization
at different doping values, for both with
 and without external magnetic fields,
and found that our M values
are qualitatively in
agreement with experimental results \cite{urushibara}.
 Furthermore our magnetoresistance values
also compare favorably with experimental ones \cite{urushibara}.

  Our starting total Hamiltonian is given by
 \begin{equation} 
 H_{T}= H(t)+H_{sp}+H_{ph}
\label{htot}
 \end{equation} 
where
 \begin{equation} 
 H(t)=t 
\sum_{\langle i j \rangle , \sigma}
  c _{i , \sigma}^{\dagger}  
  c _{j , \sigma} ,  
\label{ht}
 \end{equation} 
 \begin{equation} 
H_{ph}=\sum_{\vec{q}} \omega_{\vec{q}}
a_{\vec{q}}^{\dagger}
a_{\vec{q}} 
+ \sum_{j, \vec{q}, \sigma}
n _{j }^{\sigma} 
e^{i \vec{q} \cdot  \vec{R}_{j}}
M_{\vec{q}}
( a_{\vec{q}} +
a_{-\vec{q}}^{\dagger}) ,
\label{hph}
 \end{equation} 
and
 \begin{eqnarray*} 
H_{sp}= \sum_{\langle i j \rangle} J_{ij}
  \vec{S}_{i} \cdot \vec{S}_{j}
+ K_H \sum_{i} \vec{\sigma}_{i} \cdot \vec{S}_{i}
+ U \sum_{j, \sigma} 
n _{j }^{\sigma} 
n _{j }^{-\sigma} . 
 \end{eqnarray*} 
In the above equations 
$  c _{j , \sigma} ~ (a_{\vec{q}})$ is the
 hole (phonon) destruction operator, $t$
is the hopping integral, 
$ \langle i j \rangle$ corresponds to nearest neighbors,
$\omega_{\vec{q}}$ is the optical phonon frequency ($\hbar = 1$), $M_{q}$
is the hole-phonon coupling, $J_{ij}$ is the strength of the spin
coupling between neighboring localized  (S=3/2) spins,
$K_H$ gives the Hund's coupling between localized
 spins and
itinerant hole ($\sigma =1/2$) spin , $U$ is the strength of the same
 site repulsion, and
$n _{j }^{\sigma} =  c _{j , \sigma}^{\dagger} c _{j , \sigma}$. 
Furthermore the $H_{ph}$ part corresponds to
 assuming
a single orbital per site which on account of Jahn-Teller
splitting may perhaps be justified.

To study transport we use double exchange modification
and take the total Hamiltonian to be
 \begin{eqnarray} 
H_{T}^{tr}=
 t_{DE}
\sum _{\langle i j \rangle} 
  c _{i }^{\dagger}  
  c _{j } 
  && 
+\sum_{\vec{q}} \omega_{\vec{q}}
a_{\vec{q}}^{\dagger}
a_{\vec{q}} 
 \nonumber \\                           
  && 
+ \sum_{j, \vec{q}}
  c _{j }^{\dagger} 
  c _{j } 
e^{i \vec{q} \cdot  \vec{R}_{j}}
M_{\vec{q}}
( a_{\vec{q}} +
a_{-\vec{q}}^{\dagger}) ,
\label{HTDE}
 \end{eqnarray} 
where
$ t_{DE}= t \sqrt{(1+M^2/M_{S}^2)/2}$,
 and $M_{S}$ is the saturated
magnetization.
Now the mobility is given by the Einstein relation $\mu = q_e D \beta$
where $q_e$ is the electronic charge, $D$ the diffusivity, and $\beta
=1/k_{B} T$.
Including 
finite band width
corrections, as done variationally by Gosar \cite{gosar}, to calculate
the hopping-regime diffusivity $D_{hop}=a^2/(6 \tau )$
we obtain
 the scattering lifetime $\tau$ 
  for a {\it narrow} phononic band 
to be
 \begin{equation} 
1/(6 {\tau})=F_{[A,NA]}
 \exp{[-2 \theta \tanh (\beta \omega_{0}/4)]} ,
\label{tau}
 \end{equation} 
 where $a$ is the lattice constant,
$\omega _{0}$ is the Debye frequency, $\theta \equiv \gamma ^2 \left [
1 - \frac{(z+1) t_{DE} ^2}{2 \gamma ^4 \omega_{0}^2 } \right ] $ with
$z$ being the coordination number and, for $N$ lattice sites, 
$\gamma \omega _{0} = M_{\vec{q}} N^{1/2}$.
In the present analysis the small-but-sizeable polaron condition is
$\frac{(z+1) t_{DE} ^2}{2 \gamma ^4 \omega_{0}^2 } < 1 $. 
In Eq.\ (\ref{tau}), $F_{A}$ and $F_{NA}$ are the prefactors
for adiabatic and non-adiabatic cases with
 \begin{equation} 
F_{[A]} \equiv \frac{\omega_{0}}{2 \pi} \frac{[1+(M/M_{S})^2]}{2} ,
\label{FA}
 \end{equation} 
and
 \begin{equation} 
F_{[NA]} \equiv  \frac{t_{DE} ^2 \sqrt{\pi \sinh
( \beta \omega _{0} /2)}
 }{\omega_{0} \sqrt{\theta}},
\label{FNA}
 \end{equation} 
where the adiabatic term $ F_{[A]}$ is obtained using arguments
 similar to those 
put forth by Holstein \cite{holstein}. The crossover from
the non-adiabatic case to the adiabatic
 case occurs when $ F_{[NA]} > F_{[A]}$.
For experimentally
studied systems the adiabatic regime is of interest.
Furthermore, it should be noted 
that we need $2 \theta {\rm csch}(\beta \omega_0/2) >> 1$
for Eq.\ (\ref{tau}) to be valid
 (for a justification 
in the narrow band case $\theta =\gamma ^2$
see  Ref. \cite{holstein}).
As for the diffusivity for band conduction, it is obtained by extending
Gosar's work \cite{gosar} and calculating the polaronic band energy 
\cite{sudhakar1}
 \begin{equation} 
E _{\vec{k}}=
2 t_{DE} \exp {[- \theta \coth (\beta \omega_{0}/2 )]} 
\sum_{l}\cos (k_{l} a) .
\label{Ek}
 \end{equation} 
The above expression is similar to the result due to
Eagles \cite{eagles}.
Then the diffusivity for band conduction is given by
 \begin{equation} 
D_{band}= \langle |\vec{\nabla} E_{\vec{k}}|^2
 \tau \rangle =6 \tau a^2 \tilde{t}^{2} 
 \frac{[1+(M/M_{S})^2]}{2} ,
\label{Dband}
 \end{equation} 
where
  $\tilde{t} = t \exp {[- \theta \coth (\beta \omega_{0}/2 )]} $.
Then based on Friedman's work \cite{friedman} we take
the total mobility 
($\mu _T$)
to be the  sum of the band mobility and the hopping mobility 
and hence the total resistivity ($1/\rho = c_h q_e \mu_T$) to be given by
 \begin{eqnarray} 
\frac{4 \pi}{
 c_h q_{e}^{2} a^2 \rho} =
 \beta \omega_0  && \left [  8 \pi ^2  \frac{t^2}{\omega_{0} ^2}
 \exp {[-2 \theta 
 {\rm csch} (\frac{\beta \omega_0}{2})]} + \right . 
 \nonumber \\                           
  && 
\left . \left ( 1+  \frac{M^2}{M_{S}^2} 
 \right ) \exp{[-2 \theta \tanh 
(\frac{\beta \omega_0}{4})]} \right ],
\label{rho}
 \end{eqnarray} 
where $c_h$ is the density of holes. 
Here it should be mentioned that, 
even if $t$ and $K_H$ are of the same order of magnitude,
we can have $\tilde{t} << K_H$ so that double exchange holds.

To proceed further one needs to obtain the magnetization as a
function of temperature. To this end we consider the following
thermally averaged Hamiltonian
 \begin{equation} 
H_{mag}= 
H(\tilde{t}) + H_{sp}.
\label{hmag}
 \end{equation} 
Next we note that $|J_{ij}| << K_H$ and that
$U >> t$ $(\& K_H)$ and hence
completely project out double occupation
(see Ref. \cite{zhang} for details).
From the above Hamiltonian $H_{mag}$ it follows
 that, within a mean-field treatment, the
magnetization  in the presence of a magnetic field H is
 \begin{equation} 
S\frac{M}{M_{S}}=
- \frac {\sum_{S_z} S_{z} \exp[ -
( \Phi /2 + g  \mu_{B} H )
 S_z \beta]}
{\sum_{S_z} \exp[ -
( \Phi /2 + g  \mu_{B} H )
S_z \beta]} ,
\label{magratio}
 \end{equation} 
where
$
\Phi \equiv K_H  \left \{ n^{\uparrow}f(n^{\downarrow})
-n^{\downarrow}f(n^{\uparrow}) \right \}$,
 $f(n^{\alpha}) \equiv 1/(1-n^{\alpha})$,
and
 $n^{\alpha}$ is the probability of occupation of a site
by spin $\alpha$ hole and is given by
 \begin{eqnarray*} 
n^{\alpha} =
 \frac{1}{N} \sum _{\vec{k}} n_{\vec{k}}^{\alpha}
\left [ \epsilon _{\vec{k}}^{\alpha}
- \Psi \sigma
 -\mu \right ]
\approx
\frac{1}
{ \exp[-\beta ( 
\Psi  \sigma
 +\mu )] +1} ,
 \end{eqnarray*} 
where 
$ \Psi \equiv
(K_H S M/M_{S} 
 +  g \mu_{B} H) f(n^{\alpha})$,
$\sigma =1/2 (-1/2)$ for spin $\alpha = \uparrow (\downarrow )$ holes and
 \begin{equation} 
\epsilon_{\vec{k}}^{\alpha} =
2 \tilde{t} h(\delta ) f(n^{\alpha})
  \sum_{l}\cos (k_{l}^{\alpha}
 a) << k_{B} T_{C} .
\label{epsilonupdown}
 \end{equation} 
In the above equation $h(\delta)=1-\delta$
and $T_{C}$ is the ferromagnetic transition temperature
whose value, by treating $M$ and 
$\Delta n \equiv n^{\uparrow} -n^{\downarrow }$ as small parameters
in Eq.\ (\ref{magratio}),
 is obtained to be
 \begin{equation} 
k_{B} T_{C} = \sqrt{\frac{20 (1 - \delta ) \delta}{9 (2 - \delta )^2 }} 
K_H S |\sigma| .
\label{KBTC}
 \end{equation} 
We see that $T_{C}$ increases with increasing
$\delta$ for $0< \delta < 2/3$ and that it is independent
of both $t$ and $J_{ij}$.
Furthermore because of Eq.\ (\ref{epsilonupdown})
the values of $M$, $n^{\alpha}$, and $T_C$ are
 all independent of dimensions [see Eq.\
(\ref{magratio})].
Here it must also be mentioned that when double occupancy
is allowed
$ h(\delta ) = f(n^{\uparrow (\downarrow )}) =1 $.

Using the constraint that
$n^{\uparrow } +n^{\downarrow } = \delta $,
we can obtain $\Delta n $  and $M$
by 
solving Eq.\ (\ref{magratio}).
 In Fig.\ \ref{fig1} we have plotted the magnetization
 ratio $M/M_S$ as a function of the
reduced temperature $T/T_C$ for $\delta= 0.3$ and  $0.4$,
$g=2$, and magnetic fields $H = 0 T$ and $15 T$.
We have assumed a smaller value for the Hund's coupling
($K_H \approx 0.0858 eV$)
 than what seems to be its value based on experiments ($\sim 1 eV$)
 because
we wanted to set $T_C = 300 K$ at $\delta=0.3$.
Alternately one can also get lower $T_C$ by assuming, as suggested
 in Ref. \cite{roder},
that only a small fraction of the dopants yield mobile holes.
The values of the magnetization for $H = 15 T$ at $T_C$ are sizeable
because of the 
 tendency of the system towards a ferromagnetic phase.
Here it should also be mentioned that $\Delta n$ attains saturation
values much faster than $M$ \cite{sudhakar2}.
We have also calculated the magnetization curves with double
occupation of a site being allowed and find that the $M/M_S$ values 
for with and without
double occupation being allowed are close to each other
both in zero field and at $15 T$ 
 \cite{sudhakar2}.
 Although our magnetization curves
are qualitatively similar to the  experimental curves of Urushibara  et al.
\cite{urushibara}, the experimental $M/M_S$ values rise 
faster as $T$ is lowered.

We will now discuss 
the resistivity 
given by Eq.\ (\ref{rho}).
 The conduction goes from a hopping 
type at high temperatures to a band 
 type at low temperatures.
In Fig.\ \ref{fig2} we have shown 
 the dependence of resistivity $\rho$ on temperature
at various magnetic fields. 
The general trend of the resistivity including the drop
at the MI transition at $H = 0 T$
 is similar to the experimental results \cite{urushibara}.
On introducing a magnetic
field the system gets magnetized at temperatures higher than $T_{C}$
and thus the value of $\theta$ is smaller
 (see Eq.\ (\ref{rho})).
 Consequently the resistivity is smaller and 
 $T_{\rho^{max}}$ (the temperature at which resistivity
becomes maximum) increases \cite{sudhakar3}.

For $T \geq T_C$, 
when  $D_{band}/D_{hop} >>1$  
 the magnetoresistance 
$\Delta \rho/\rho(0)\equiv 
(\rho(H)-\rho(0))/\rho(0)$ is given by (see Eq.\ (\ref{rho}))
 \begin{equation} 
\Delta \rho/\rho(0) 
\approx 
\exp \left [ - \frac {(z+1)}{2 \gamma ^2} \frac{ t^2}{\omega_{0} ^2} 
\frac{ M^2}{ M_{S}^2}
{\rm csch}(\frac{\beta \omega_{0}}{2} ) \right ] - 1 ,
\label{deltarhoband}
 \end{equation} 
and when $D_{band}/D_{hop} <<1$ it is given by
 \begin{equation} 
\Delta \rho/\rho(0) 
\approx  \frac{\exp 
\left [ -\frac{(z+1)} {2 \gamma ^2 }
\frac{t^2}{ \omega_{0}^2}
 \frac{ M^2} { M_{S}^2} 
 \tanh(\frac{\beta \omega_{0}}{4} ) \right ] 
 }{1 + (M/M_{S})^2} - 1 .
\label{deltarhohop}
 \end{equation} 
For a fixed value of
the reduced temperature $T/T_C$,
an increase in the ratio $\mu_B H/K_H$ increases
$M/M_S$
and consequently  the magnetoresistance also increases.

Actually  $T_{\rho_{M=0}^{max}}$
(the temperature at which the resistivity  given by Eq.\ (\ref{rho}),
after taking $M=0$, attains a maximum) 
need not be equal to the ferromagnetic transition temperature $T_{C}$.
If $T_{\rho_{M=0}^{max}} < T_{C}$, by decreasing $\gamma ^2$ or
increasing $\frac{t^2}{\omega_{0}^2}$
activation energy ($\theta \omega_{0}/2$) decreases and
 $T_{\rho_{M=0}^{max}}$ can be increased \cite{sudhakar3} to be 
made equal to $ T_{C}$ and this also increases the magnetoresistance
(see Eqs.\ (\ref{rho}), (\ref{deltarhoband}), and (\ref{deltarhohop})).
For $T_{\rho_{M=0}^{max}} < T_{C}$,
the MI transition can still occur at $T_{C}$
 if $\frac{(z+1)}{2 \gamma^2} \frac{t^2}{\omega_{0}^2}$
 is sufficiently large \cite{sudhakar3} while
 if $\frac{(z+1)}{2 \gamma^2} \frac{t^2}{\omega_{0}^2}$
is very small the MI transition occurs below
$T_{\rho_{M=0}^{max}}$
as can be seen from Eq.\ (\ref{rho}).
The other case,
where
$T_{\rho_{M=0}^{max}} > T_{C}$,  corresponds
to MI transition occurring at a higher temperature than $T_C$
 and is in any case not experimentally observed \cite{sudhakar2}.

In Table \ref{tab1} we report the calculated 
values of magnetoresistance 
$-\Delta \rho/\rho(0)$ at $T_C$ and the
optimum values of $\gamma^2$ (obtained when
 $T_{\rho_{M=0}^{max}} = T_{C}$)
for doping $\delta$ equal to $0.3$ and $0.4$, Debye temperature
$T_D = 500 K$, and for various values of the dimensionless
hopping integral
$t/\omega_0$. 
We find that the magnetoresistance increases with
increasing values of $t/\omega_0$
thus showing the importance of bandwidth. Also 
 $\gamma^2_{opt}$ values increase with increasing $t/\omega_0$ 
because
 $T_{\rho_{M=0}^{max}} = T_{C}$.
Furthermore, it is mainly due to the larger values of $(M/M_S)^2$
for $\delta=0.4$ compared to those of $\delta=0.3$ that the
values of $-\Delta \rho/\rho(0)$ are larger for $\delta=0.4$.
It appears that our model can give magnetoresistance values
comparable to the experimental ones \cite{urushibara}.
In fact one can get a larger magnetoresistance by taking
a smaller $T_C$ value but keeping $\omega_0/(k_B T_C)$ 
fixed \cite{sudhakar2}.
                                     
        From Eq.\ (\ref{rho})
(or Eqs.\ (\ref{deltarhoband}) and (\ref{deltarhohop}))
 we see that for small values of $M/M_{S}$ 
the magnetoresistance (for $T \geq T_C$)
  is of the form $-\Delta \rho/\rho(0) = C (M/M_S)^2 $
where $C$ is a constant of proportionality.
We found, for the cases considered in Table 1, that
the optimum values of $\gamma ^2$ that 
 make $T_{\rho_{M=0}^{max}} = T_{C}$
 are such that $D_{band}/D_{hop} < 1$ so that
 the magnetoresistance can be qualitatively given
 by Eq.\ (\ref{deltarhohop}). 
In Eq.\ (\ref{deltarhohop}),
  close to $T_C$,  $\tanh (\beta \omega_{0}/4 ) \approx 
\beta \omega_{0}/4 $. From Eq.\ (\ref{KBTC}) 
 we see that $T_C$ increases with
the doping $\delta$ and hence the constant of proportionality $C$ 
($\propto  \frac{ (2 - \delta )} {\sqrt{(1 - \delta ) \delta}}$)
 decreases with increasing $\delta$ which agrees
with the findings of Ref. \cite{urushibara}.  
Furthermore the coefficient also increases with increasing values of
$ \frac{(z+1) t^2 }{ \gamma ^2 T_C \omega_{0} } $.
We have calculated values of $C$ by
treating $M/M_S$ as a small parameter in the 
exact expression for $-\Delta \rho/\rho(0)$ at $T=T_C$. 
 When $t/\omega_0 =4 (8)$ and $\gamma^2=9.5 (15.0)$,
for $\delta=0.3$ we get 
 $C\approx 3.8 (8.5)$ while  for $\delta=0.4$
we obtain $C \approx 3.1 (6.4)$. 
Our calculated values of $C$ 
 are larger than those reported in Ref. \cite{urushibara}. 
 Past attention \cite{inoue,furukawa} has focused at dependence of $C$
on the ratio $K_H/t$ in Kondo lattice type models that ignored
 electron-phonon coupling.
 While Inoue and Maekawa \cite{inoue}  
for $K_H \rightarrow \infty$ obtained
$C=7/4$, Furukawa \cite{furukawa} found 
that the value of $C$ increased with
increasing values of $K_H/t$ and that at
 larger values of $K_H/t$ the value of $C$
decreases with increasing doping. 

In conclusion we say that both double exchange
 and finite band-width corrections
are important to understand CMR.
 In our picture, adiabatic 
 small-but-sizeable magnetic polarons  are involved in
activated transport  at high temperatures
and metal-like conduction at low temperatures.
 At the MI transition, the band-like
motion of the carriers also produces a paramagnetic-ferromagnetic
transition due to strong Hund's coupling between
itinerant and localized spins. 
Studying the transport behavior at low temperatures, including
a Fermi liquid analysis, is left for future.
The effect of including both $d_{3 z^2 - r^2}$ and $d_{x^2 - y^2}$
orbitals also needs to be investigated for our model.
 Lastly we note that as the
system's temperature is lowered below $T_C$ the
magnetization increases 
and consequently
the activation 
 energy 
($\theta \omega_0/2$)
 decreases and the
polarons tend towards large polaronic behavior.

\vspace{10cm}
\narrowtext                                                
\begin{figure}
\caption{Plot of the magnetization ratio $M/M_{S}$ versus
the reduced temperature $T/T_{C}$
when  no double occupation is allowed,
 doping $\delta=0.3$ (and $ 0.4$), 
Hund's coupling
 $K_H \approx 0.0858 eV$, 
 and magnetic fields 
 $H = 0 T$ and $H = 15 T$.} 
\label{fig1}
\end{figure}

\begin{figure}
\caption{Plot of the resistivity $\rho$ in units of
 $4 \pi /(c_h q_{e}^2 a^2)$
versus temperature $T$
in 3 dimensions 
when no double occupation is allowed, 
 $\delta=0.3$, 
 $K_H\approx 0.0858 eV$, 
 dimensionless hopping integral $t/ \omega_0 = 6$,
 optimum $\gamma^2 = 12.2$,
Debye temperature $T_D =500 K$, and for the following magnetic
fields:
(i)  $H = 0 T$; 
(ii) $H = 15 T$; 
(iii) $H = 30 T$; and
(iv) $H = 45 T$.} 
\label{fig2}
\end{figure}

\begin{table}
\caption{Calculated values of 
 the magnetoresistance $-\Delta \rho/\rho (0)$ at $T_C$ and 
the optimum $\gamma^2$ 
 for various values
of $t/ \omega_0$, 
$T_D =500 K$, $\delta= 0.3$ and $ 0.4$,
magnetic field $H=15 T$, 
 $K_H\approx 0.0858 eV$, and
 $T_{\rho_{M=0}^{max}}= T_C$.}
\label{tab1}
\end{table}

\begin{tabular}{l|rc|rc}   \hline\hline
  & \multicolumn{2}{c|}{$\delta = 0.3$ ; $T_C = 300 K$}& 
\multicolumn{2}{c}{$\delta = 0.4$ ; $T_C \approx 341 K$} \\ \cline{2-5}    
$ t/ \omega_0$  & $\gamma^2_{opt}$   & $ \frac{-\Delta \rho}{\rho (0)} $ 
  
  &  $\gamma^2_{opt}$   & $  \frac{-\Delta \rho}{\rho (0)} 
$ 
   \\  \hline
4   &   9.5   & 35\% & 8.4 & 42\% \\ \hline
5   &   10.8  & 44\% & 9.8 & 50\%  \\ \hline
6   &   12.2  & 51\% & 11.2 & 58\% \\ \hline
7   &   13.6  & 58\% & 12.5 & 68\% \\ \hline
8   &   15.0  & 64\% & 13.9 & 74\% \\ 
 \hline\hline
\end{tabular}

\end{multicols} 

\begin{references}
\bibitem{sudhakar1}   
We get the polaronic band energy  by replacing
$ \gamma_{\lambda}/\sqrt{N} $ with $z_{\lambda}$ in the 
second term in Eq.\ (36) of Ref. \cite{gosar}.
\bibitem{sudhakar3}   
For $\beta \omega_{0}/2 < 1$, on taking
${\rm csch}(\beta \omega_0/2) \approx 
2/(\beta \omega_{0})$  and 
$\tanh (\beta \omega_0/4) \approx 
\beta \omega_{0}/4$, 
if $\theta \beta \omega_{0}/2 > 1$ 
the following can be shown: 
(i) $T_{\rho_{M}^{max}}$ increases as $\gamma ^2$ decreases
(or $t^2/\omega_{0}^2$ increases); and (ii) for fixed
$\gamma^2 $ and $t^2/\omega_{0}^2$
 and for large enough
 $\frac{(z+1)}{2 \gamma^2} \frac{t^2}{\omega_{0}^2}$,
 $T_{\rho^{max}}$ increases as $M$ increases.
\end{references}
  
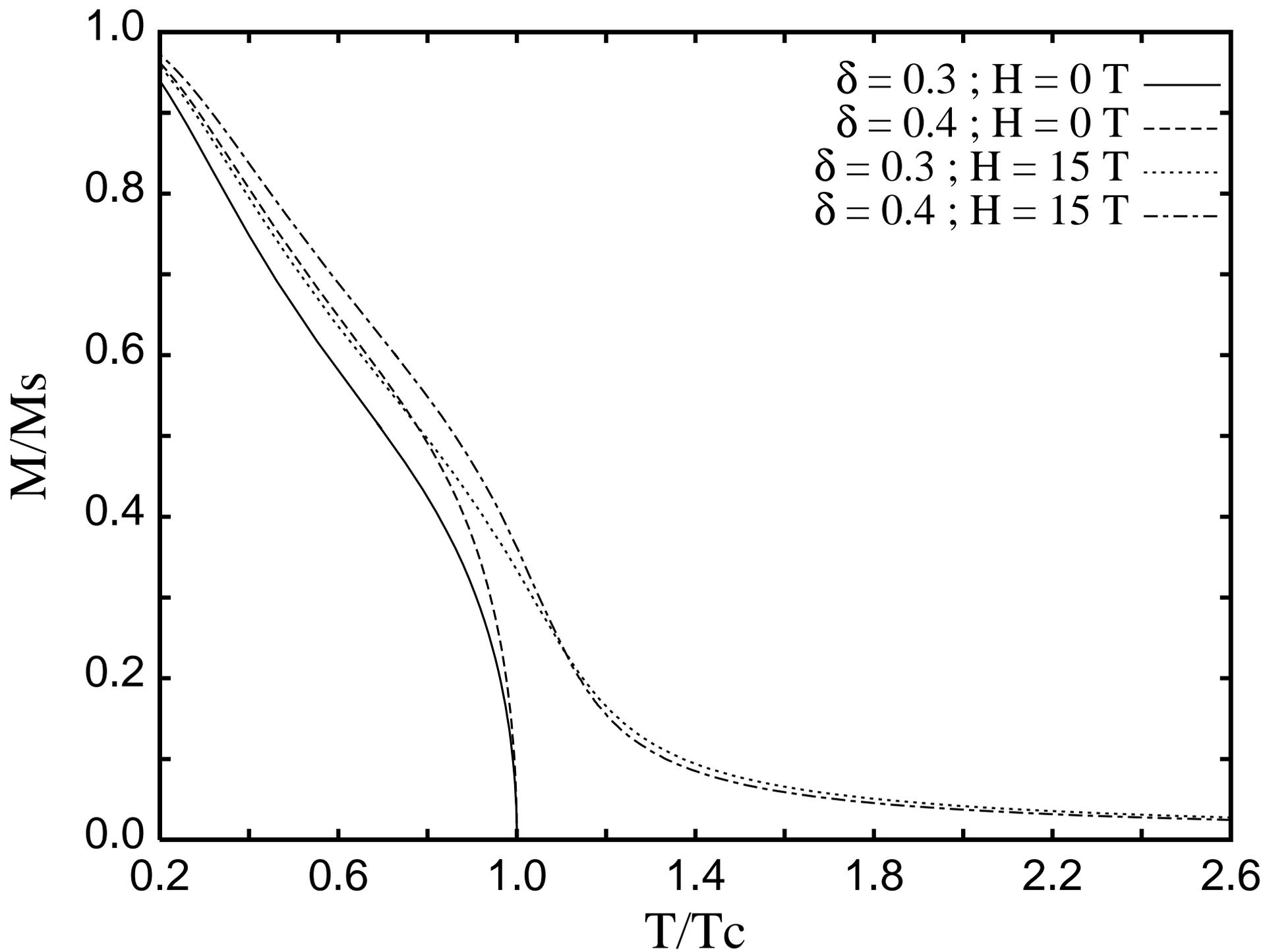
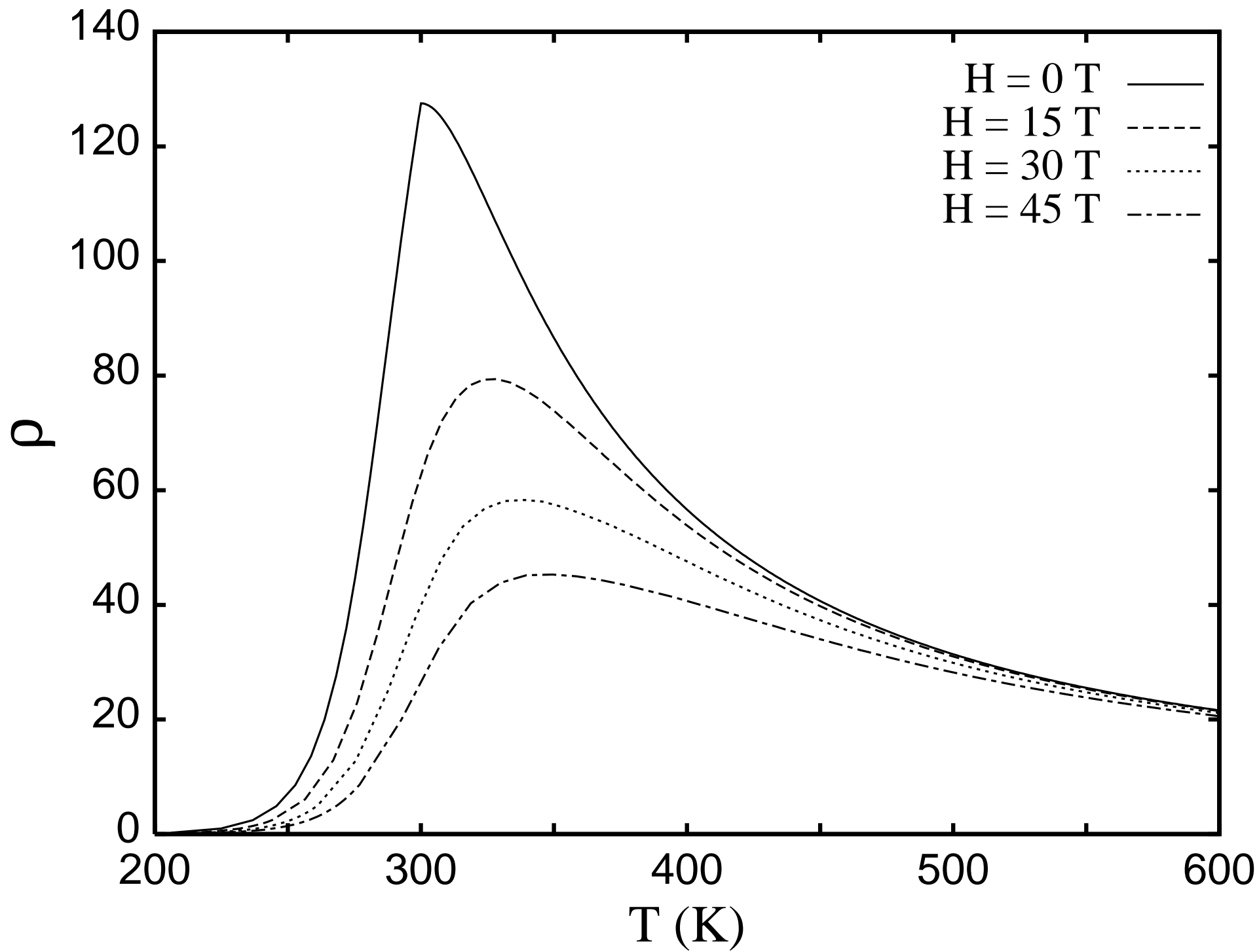
\end{document}